\documentclass[a4paper]{jpconf}
\usepackage{graphicx}
\usepackage[german,english]{babel}
\usepackage[nospace]{cite}
\usepackage{upgreek}
\begin{document}
\title{Towards Acoustic Detection of UHE Neutrinos in the
  Mediterranean Sea \--- \\ The AMADEUS Project in ANTARES\footnote{This
    work is supported by the German BMBF Grant No.~05~CN5WE1/7.}}

\author{K~Graf, G~Anton, J~H{\"o}{\ss}l, A~Kappes, U~F~Katz, R~Lahmann,
  C~Naumann and K~Salomon} \address{Physikalisches Institut,
  Friedrich-Alexander-Universit{\"a}t Erlangen-N{\"u}rnberg,
  Erwin-Rommel-Stra{\ss}e 1, D-91058 Erlangen, Germany}
\ead{graf@physik.uni-erlangen.de}

\begin{abstract}
  The acoustic detection method is a promising option for future
  neutrino telescopes operating in the ultra-high energy regime. It
  utilises the effect that a cascade evolving from a
  neutrino interaction generates a sound wave, and is applicable in
  different target materials like water, ice and salt.  Described here
  are the developments in and the plans for the research on acoustic
  particle detection in water performed by the ANTARES group at the
  University of Erlangen within the framework of the ANTARES
  experiment \cite{antares_proposal} in the Mediterranean Sea. A set
  of acoustic sensors will be integrated into this optical neutrino
  telescope to test acoustic particle detection methods and perform
  background studies.
\end{abstract}
\section{Introduction}

Towards the detection of neutrinos with energies exceeding 100\,PeV,
the use of acoustic pressure waves produced in neutrino-induced
cascades is a promising approach which is investigated by several
running and planned projects, see e.g.\
\cite{vandenbroucke_apj,proc_arena}. The acoustic wave originates from
the heating of the medium in the vicinity of the evolving cascade \---
a mechanism which is described by the thermo-acoustic model
\cite{askarian_nim}.  Once generated, the sound wave propagates in a
flat disk shape perpendicular to the main axis of the cascade. In a
sensor, the resulting signal is bipolar in time and has its main
frequency components in the range from 1 to 100\,kHz.  The absorption
length of sound in sea water at the peak spectral density around
20\,kHz is on the order of a kilometre \cite{fisher_jasa}. This would
make it possible to instrument efficient acoustic detectors with 200
sensor clusters per km${^3}$ \cite{karg_phd,perkin_arena05}.  Given
the expected low flux of neutrinos with energies in excess of 100\,PeV, a
potential acoustic neutrino telescope must not only have large
dimensions, but also has to be operated basically background-free.

To investigate the feasibility of building a detector in the deep-sea
based on this method, it is therefore necessary to understand the
acoustic background conditions and characteristics of transient noise
sources at the site in detail. Especially the knowledge of the rate
and correlation length of acoustic background events with
neutrino-like signature is a prerequisite for the estimation of the
detector sensitivity.  Thus the aim of the project {\it AMADEUS}
(\underline{A}NTARES \underline{M}odules for \underline{A}coustic
\underline{De}tection \underline{U}nder the \underline{S}ea),
described in the following, is to measure the acoustic conditions of
the deep-sea environment at the ANTARES site with a dedicated array of
custom-designed acoustic sensors at different distances over a
long time scale.

For these studies several additional basic detector elements ({\it
  storeys}, cf.~Sec.~\ref{cap_antares}), equipped with acoustic
sensors, will be installed in the ANTARES neutrino telescope. On these
storeys the acoustic sensors will substitute the optical sensors used
for Cherenkov detection of neutrinos.

\section{The ANTARES detector}\label{cap_antares}

\begin{figure}[h]
\centering
\includegraphics[width=14cm]{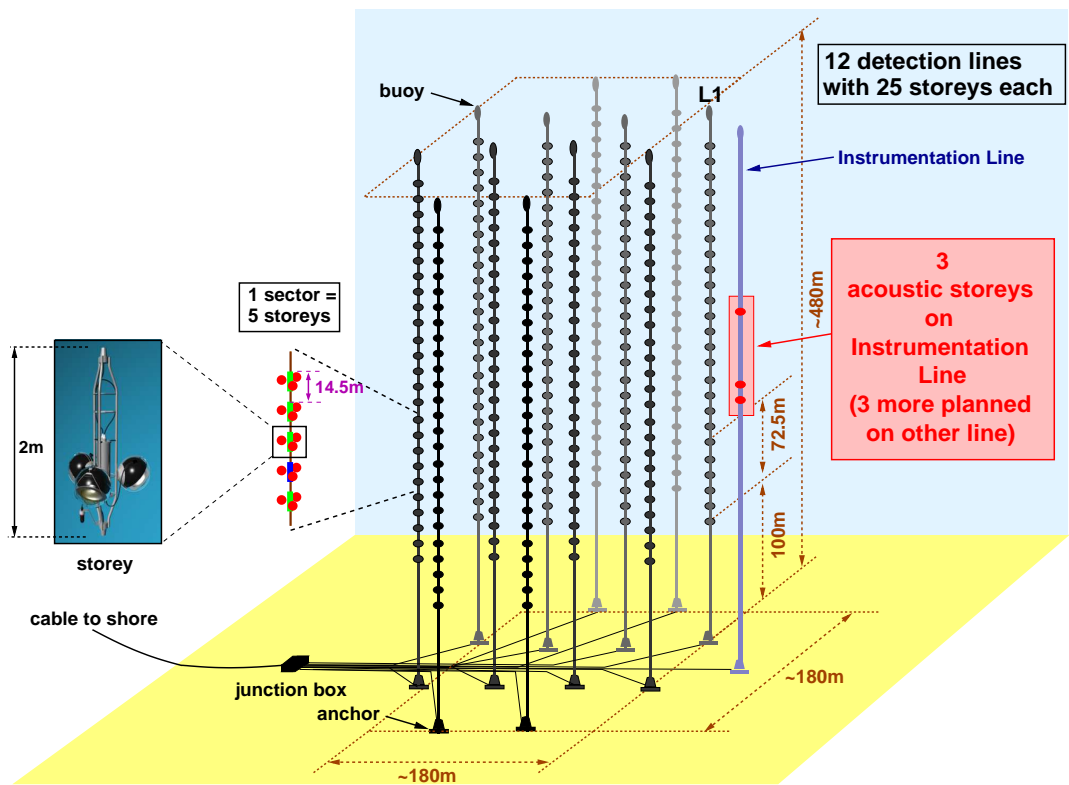}
\caption{Sketch of the ANTARES detector with the acoustic addition
  AMADEUS. For further description see
  text.\label{fig_antares_scheme}}
\end{figure}

Figure \ref{fig_antares_scheme} shows a sketch of the complete ANTARES
detector with the acoustic AMADEUS module, which is further
described in Sec.~\ref{sec_acoustics}.
\\The ANTARES detector \cite{antares_proposal} is currently
constructed in the Mediterranean Sea in a water depth of up to
2500\,m. Its completion is planned for 2007; the final Cherenkov
neutrino telescope will consist of 12 vertical structures ({\it
  detection lines}). The Instrumentation Line ({\it IL}) \--- an extra
13th line \--- will be equipped with sensors to monitor environmental
parameters in the deep sea and with devices to calibrate the detector.
Its installation is planned for mid-2007. Each detection line has a
total height of 460\,m and comprises 25 {\it storeys} (cf.~left insert
in Fig.~\ref{fig_antares_scheme}) spaced evenly within the
instrumented height of 350\,m.  The storey is the basic detection
element and consists of three Optical Modules ({\it OMs}) (optical
sensors in a pressure-resistant glass housing), a Local Control Module
({\it LCM}) (for data acquisition-, control- and monitoring hardware)
and miscellaneous auxiliary devices on a mechanical support frame. The
12 detection lines will cover a total area of approx.\
180\,$\times\,$180\,m{$^2$} on the sea-floor.  The detector is
connected to the on-shore control room via deep-sea cables providing
electrical power and data transmission. At the writing of these
proceedings, two detection lines and a progenitor of the IL have been
installed and are operated successfully.

\section{The Acoustic Setup AMADEUS}\label{sec_acoustics}

Three storeys on the IL will be equipped with six acoustic sensors
each. The vertical spacing for these storeys will be approx.\ 15\,m
and 100\,m.  Together with the sensor spacing of approx.\ 1\,m within
the storey, this will provide three different length scales for the
investigation of acoustic background sources. Additional three
acoustic storeys are planned on one further detector line at a
horizontal distance exceeding 100\,m.  For the integration of acoustic
sensors into the ANTARES experiment the data acquisition ({\it DAQ})
system \cite{antares_daq} and some mechanical structures have to be
modified.  This is done under the premise of preventing any
interference with the optical data taking. To optimise
resources and to make use of the well-tested, existing system wherever
feasible, as little changes as possible to the ANTARES design are
targeted.
\subsection{Acoustic Sensors}

\begin{figure}[h]
\centering
\includegraphics[height=5.5cm]{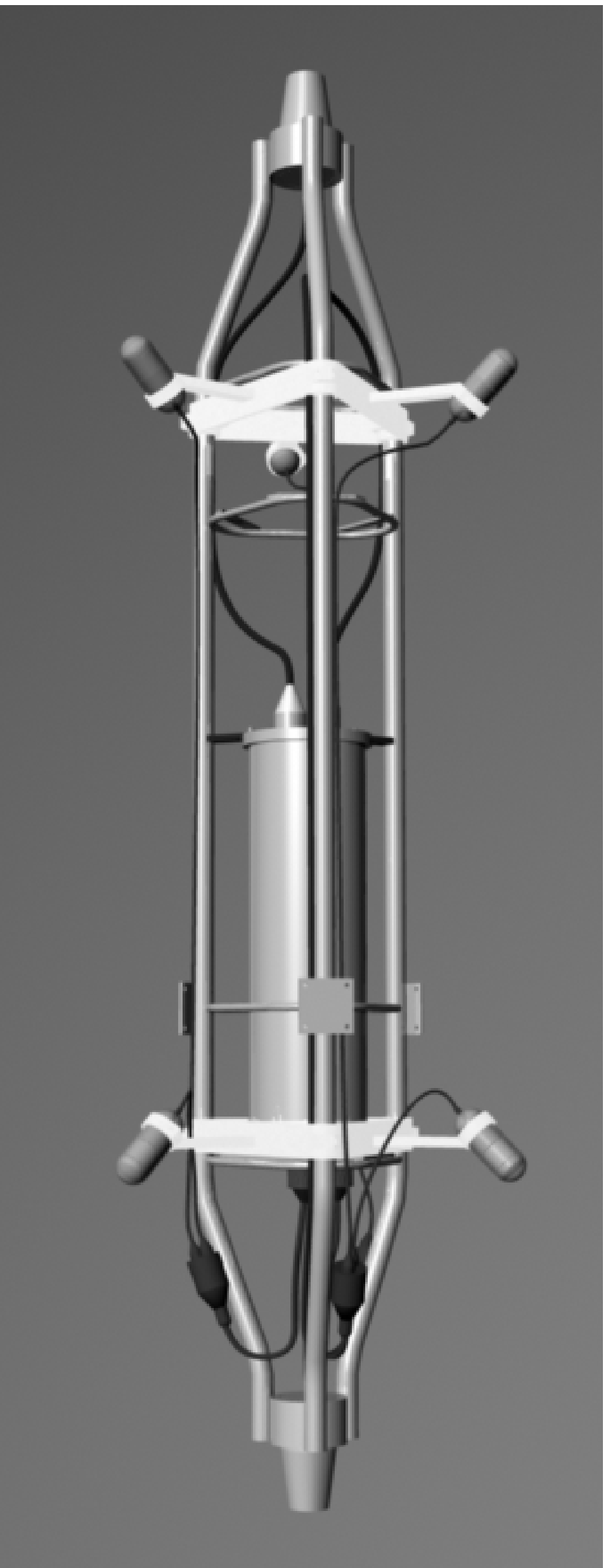}\hspace{2pc}
\includegraphics[height=5.5cm]{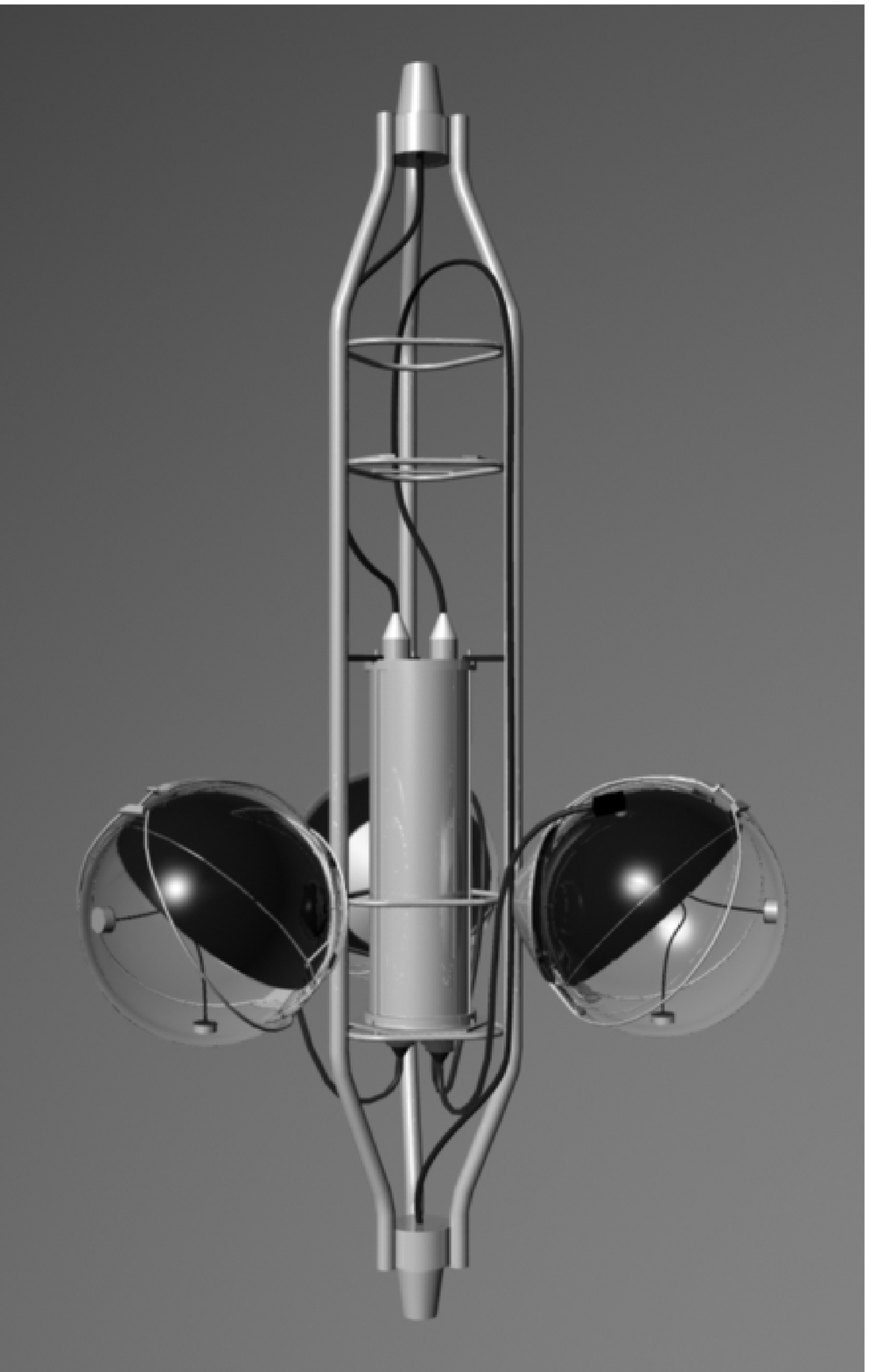}\hspace{2pc}
\begin{minipage}[b]{8cm}
  \caption{Artist's views of acoustic storeys with six hydrophones
    (left) and three acoustic modules housing two piezo sensors each
    (right).\label{fig_antares_storey_acou}}
\end{minipage}
\end{figure}

Major changes affect the storey, where the optical sensors are
replaced by acoustic ones: hydrophones or so-called acoustic modules
({\it AMs}) \cite{naumann_arena05}. Artist's views of the resulting
acoustic storeys are shown in Fig.~\ref{fig_antares_storey_acou}.
\\The acoustic sensors are based on piezo ceramics that convert
pressure waves into voltage signals, which are amplified for read-out
\cite{hoessl_app}. In the case of the hydrophones, the ceramics and
amplifiers are coated in polymer plastics. For the AMs they are glued
to the inside of a water-tight sphere. The three storeys on the IL
will house hydrophones only, whereas at least one of the storeys
planned additionally will house the non-conventional but promising
design of AMs \cite{naumann_arena05}. All acoustic sensors are
custom-designed. They are tuned to be sensitive over the whole
frequency range of interest (around -145\,dB\,re.~1V/$\upmu$Pa) and to
have a low noise level.

\subsection{Acoustic Data Acquisition}

The acoustic signals are preprocessed and digitised within the LCM on
each storey by custom-designed electronics boards \--- the acoustics
digitisation boards ({\it AcouADC-boards}).  These boards are
integrated into the ANTARES DAQ system, which provides
nanosecond-scale timing resolution, positioning of the storeys and
transmission of the data to the on-shore control room.  There are a
total of three such boards per storey receiving the data of two
sensors each.

\subsection{The AcouADC-Board and System Performance}

The AcouADC-board consists of an analogue and a digital part. The
analogue part amplifies and filters the acoustic signals coming from
the sensors. The system has low noise and is designed to be \---
together with the sensors \--- sensitive to the acoustic background of
the deep sea over a wide frequency range.  A bandpass filter with
cut-off frequencies of approx.\ 1 and 100\,kHz is integrated to avoid
the trailing edge of the low frequency noise \cite{urick} and aliasing
effects from frequencies above the Nyquist frequency of the
digitisation.

The acoustic data from the analogue part is digitised with a 16-bit
resolution and a maximum sampling rate of 500\,k\,Samples per second
({\it kSPS}) and then processed in the digital part of the board.
There the data can be down-sampled to reduce data traffic and is read
out by the ANTARES DAQ system which handles the transmission to the
control room.  The maximum sample rate per storey is bandwidth-limited
to an average of 1.25\,MSPS. Thus the data of all six sensors can be
transmitted with a sampling of 200\,kSPS, or alternatively the data of
two sensors at full rate.

The whole data-taking chain is calibrated allowing for precise
reconstruction of the acoustic signal from the recorded one within the
sensitive frequency range of the set up. The dynamic range achieved is
from the order of 1\,mPa to the order of 10\,Pa (rms over the frequency range
from 1 to 100\,kHz). This allows for studying both, the acoustic
background in the deep sea under all prevailing conditions \cite{urick}
and transient sounds with neutrino-like signatures.

\subsection{On-shore Control Room}
The DAQ system will be controlled from the on-shore control room. From
there, settings for the analogue part and the data processing in the
digital part can be adjusted. An update of the programming code of the
AcouADC-board is also possible in situ.  A dedicated PC-cluster will
be set up to process and store the acoustic data arriving from the
storeys. On this cluster different data filtering schemes and
triggers will be implemented, and an adequate amount of raw data will be
stored.

\section{Summary and Outlook}

We described AMADEUS, a project to investigate the feasibility of a
future neutrino detector using the acoustic detection method in water.
With this project, a dedicated array of acoustic sensors will be
installed in the deep sea at the ANTARES site in 2007. Long-term
studies of the acoustic background noise and signals in this
environment will be performed.  The main goal is to measure the rate
of correlated neutrino-like background events and their correlation
length, which is decisive for assessing the sensitivity of a future
acoustic detector for ultra-high-energy neutrinos.

\end{document}